\documentclass[reprint,showpacs,preprintnumbers,amsmath,amssymb,aps,prc,superscriptaddress]{revtex4-1}

\usepackage{graphicx}
\usepackage{dcolumn}
\usepackage{bm}

\begin{document}


\title{Determination of $^{141}$Pr($\alpha$,n)$^{144}$Pm cross sections at energies of relevance for the astrophysical $p$ process using the $\gamma \gamma$ coincidence method}

\author{A.~Sauerwein}
\email[]{sauerwein@ikp.uni-koeln.de}
\affiliation{Institut f\"ur Kernphysik, Universit\"at zu K\"oln, Z\"ulpicher Stra\ss{}e 77, 50937 K\"oln, Germany}

\author{H.W.~Becker}
\affiliation{Dynamitron Tandem Labor des RUBION, Ruhr-Universit\"at Bochum, Universit\"atsstra\ss{}e 150, 44780 Bochum, Germany}

\author{H.~Dombrowski}
\affiliation{Physikalisch-Technische Bundesanstalt (PTB), Bundesallee 100, 38116 Braunschweig, Germany}

\author{M.~Elvers}
\affiliation{Institut f\"ur Kernphysik, Universit\"at zu K\"oln, Z\"ulpicher Stra\ss{}e 77, 50937 K\"oln, Germany}

\author{J.~Endres}
\affiliation{Institut f\"ur Kernphysik, Universit\"at zu K\"oln, Z\"ulpicher Stra\ss{}e 77, 50937 K\"oln, Germany}

\author{U.~Giesen}
\affiliation{Physikalisch-Technische Bundesanstalt (PTB), Bundesallee 100, 38116 Braunschweig, Germany}

\author{J.~Hasper}
\affiliation{Institut f\"ur Kernphysik, Universit\"at zu K\"oln, Z\"ulpicher Stra\ss{}e 77, 50937 K\"oln, Germany}

\author{A.~Hennig}
\affiliation{Institut f\"ur Kernphysik, Universit\"at zu K\"oln, Z\"ulpicher Stra\ss{}e 77, 50937 K\"oln, Germany}

\author{L.~Netterdon}
\affiliation{Institut f\"ur Kernphysik, Universit\"at zu K\"oln, Z\"ulpicher Stra\ss{}e 77, 50937 K\"oln, Germany}

\author{T.~Rauscher}
\affiliation{Departement f\"ur Physik, Universit\"at Basel, Klingelbergstra\ss{}e 82, 4056 Basel, Switzerland}

\author{D.~Rogalla}
\affiliation{Dynamitron Tandem Labor des RUBION, Ruhr-Universit\"at Bochum, Universit\"atsstra\ss{}e 150, 44780 Bochum, Germany}

\author{K.O.~Zell}
\affiliation{Institut f\"ur Kernphysik, Universit\"at zu K\"oln, Z\"ulpicher Stra\ss{}e 77, 50937 K\"oln, Germany}

\author{A.~Zilges}
\affiliation{Institut f\"ur Kernphysik, Universit\"at zu K\"oln, Z\"ulpicher Stra\ss{}e 77, 50937 K\"oln, Germany}

\date{\today}

\begin{abstract}
The reaction $^{141}$Pr($\alpha$,n)$^{144}$Pm was investigated between $E_\alpha=$ 11 MeV and 15 MeV with the activation method using the $\gamma \gamma$ coincidence method with a segmented clover-type
high-purity Germanium (HPGe) detector. Measurements with four other HPGe detectors were additionally made. The comparison proves that the $\gamma \gamma$ coincidence method is an excellent tool to
investigate cross sections down to the microbarn range.
The ($\alpha$,n) reaction at low energy is especially suited to test $\alpha$+nucleus optical-model potentials for application in the astrophysical $p$ process. The experimentally determined cross sections were compared to Hauser-Feshbach statistical model calculations using different optical potentials and generally an unsatisfactory reproduction of the data was found. A local potential was constructed to improve the description of the data. The consequences of applying the same potential to calculate astrophysical ($\gamma$,$\alpha$) rates for $^{145}$Pm and $^{148}$Gd were explored. In summary, the data and further results underline the problems in global predictions of $\alpha$+nucleus optical potentials at astrophysically relevant energies.
\end{abstract}

\maketitle

\section{\label{sec:introduction}Introduction}
According to the current understanding most nuclei heavier than iron are synthesized by neutron-capture reactions via the $s$ and $r$ processes \cite{b2fh,agw,ctt,wall97,arngorr,kapgall}. About 35 proton-rich nuclei in this mass region, however, are bypassed by these processes. These nuclei are called $p$ nuclei. The origin of the $p$ nuclei is not completely understood and contributions of different independent processes for their production are under discussion. The original suggestion to produce proton-rich nuclides in a $p$ process, (i.e., by proton capture reactions in the H-rich envelope of type II supernovae \cite{b2fh}) was later shown to be unfeasible \cite{autru,arngorp}.

It was found, however, that the O/Ne layers of a massive star are sufficiently heated during the passage of the explosive shock wave of a core-collapse-induced supernova to allow partial photodisintegration of the $s$- and $r$-process nuclei previously present in the stellar plasma \cite{arn,woohow,rayet95}.
The photodisintegration reactions produce $p$ nuclei through sequences of ($\gamma$,n),
($\gamma$,p), and ($\gamma$,$\alpha$) reactions at plasma temperatures in the range of $2\leq T\leq 3$ GK. This so-called $\gamma$ process is the currently accepted nucleosynthesis mechanism to explain the majority of $p$ nuclei. Among the few exceptions are $^{138}$La and $^{180m}$Ta, both of which have very low abundances in the solar system. They could be produced through neutrino reactions with neutrinos emitted
by the nascent neutron star emerging from the core collapse (the $\nu$ process) \cite{neutrino}. Some $p$ nuclei ($^{164}$Er and $^{152}$Gd \cite{arl99}, $^{113}$In and $^{115}$Sn \cite{nemeth}) may also receive stronger contributions from the $s$ and $r$ process than previously estimated and may not require a large $\gamma$-process production.

In addition to the above exceptions, two mass regions have remained problematic in explaining the production of $p$ nuclei by the $\gamma$ process in core-collapse supernovae: the lightest $p$ nuclei with mass numbers $A<100$ and those in an intermediate region at $150\leq A \leq165$ are under produced \cite{arngorp,woohow,rayet95,rhhw02,hegXX}. It is not yet fully understood to which extent the deficiencies found are due to the astrophysical modeling or the nuclear physics input (reaction rates) \cite{raupp11}. For example, recent simulations \cite{travaglio} found that light $p$ nuclei are produced in sufficient amounts in a $\gamma$ process occurring in the thermonuclear explosion of a white dwarf (type Ia supernova), in contrast to earlier simulations finding no such production \cite{howmeywoo,howmey,kusa05,kusa11}. Regardless of the site, such calculations also implement a $\gamma$ process and require a sound determination of the astrophysical reaction rates appearing from nuclear physics.

In total, the $\gamma$ process involves an extensive reaction network consisting of about twenty thousand reactions on approximately two
thousand nuclei. Most of these nuclei are unstable and therefore are not directly accessible for experiments. Due to the astrophysically relevant low interaction energies, data is also scarce in the relevant energy range for stable nuclei. In consequence, reaction network calculations for the $\gamma$ process are based almost completely on theoretically predicted reaction rates stemming from Hauser-Feshbach statistical model calculations \cite{Haus52,NONSMOKER}. The accuracy of the predictions mainly depends on the adopted nuclear models for optical-model potentials, $\gamma$-strength functions,
and nuclear-level densities.

The low-energy $\alpha$+nucleus optical potential determines deflections in the photodisintegration path at intermediate and high masses and therefore impacts the calculated $p$ abundances also in the problematic region of $150\leq A \leq165$ \cite{raudeflect,rappgamma}. The few experimental studies available close to the astrophysical energy region in this mass range revealed a systematic overprediction of the ($\alpha$,$\gamma$) cross sections with the widely used potential of \cite{McFa66} (see, e.g., \cite{tm169,yalcin,Hari05,gyurky,151eu,rapp02,rapp08} and references therein). The predictions are mostly factors of 2 to 3 above the data, with the exception of \cite{somo98} where the measured $^{144}$Sm($\alpha$,$\gamma$) $S$ factors were found to be lower by more than an order of magnitude than the standard prediction at astrophysical energies. So far, the data at low energy are still too scarce to allow the construction of a global optical potential suited to predict astrophysical reaction rates further off stability, as required by the $\gamma$ process.

In the present work, the reaction $^{141}$Pr($\alpha$,n)$^{144}$Pm was studied with the activation technique (for details, see \cite{rapp02}) at the cyclotron of the Physikalisch-Technische Bundesanstalt (PTB) in Braunschweig, Germany \cite{Bred80} to improve the experimental situation on the $\alpha$+nucleus optical-model potential. The case of $^{141}$Pr+$\alpha$ is especially interesting because $^{141}$Pr is close to $^{144}$Sm and is also neutron-magic. It is important to see whether similar problems in the prediction of cross sections and astrophysical $S$ factors arise for reactions on this nucleus as those found for $^{144}$Sm. Sensitivity studies show that, in the case of $^{141}$Pr, the ($\alpha$,n) reaction is better suited to improve the $\alpha$+nucleus optical-model potential compared to the ($\alpha$,$\gamma$) reaction (see Sec.\ \ref{sec:sensitivities}). The ($\alpha$,n) reaction was investigated at laboratory $\alpha$ energies between 11.0 MeV and 15.0 MeV.

The sensitivities of the predicted reaction cross sections to different nuclear input are discussed in Sec.\ \ref{sec:sensitivities}.
After presenting the experimental method in Sec.\ \ref{sec:experiment} the data analysis is explained in Sec.\ \ref{sec:data analysis}. The deduced cross
sections are compared to theoretical predictions from Hauser-Feshbach statistical model calculations in Sec.\ \ref{sec:results}.

\section{\label{sec:sensitivities}Sensitivity Studies}

\begin{figure}
\begin{center}
\includegraphics[width=\columnwidth]{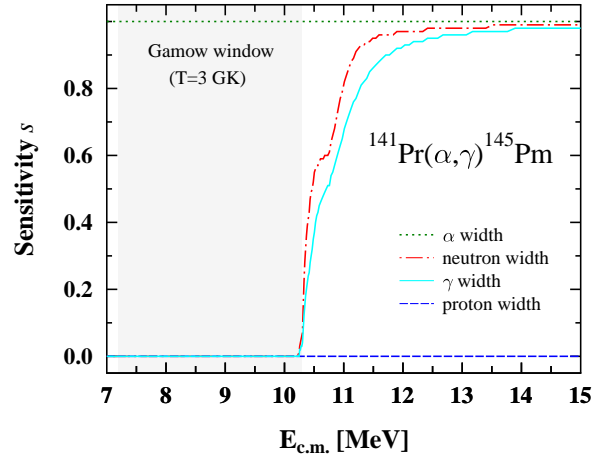}
\end{center}
\caption{\label{fig:Sensitivity_ag} Sensitivities $s$ of the $^{141}$Pr($\alpha$,$\gamma$) laboratory cross sections when varying neutron, proton, $\alpha$, and $\gamma$ widths separately by a factor of 2. The astrophysically relevant energy range for $T=3$ GK is marked by the shaded area.
Within this energy window the cross sections are almost exclusively sensitive to the $\alpha$ width, whereas at energies
measurable with the activation method within a reasonable time (well above 10 MeV), the cross section prediction
is sensitive to additional nuclear parameters as well. }
\end{figure}

\begin{figure}
\begin{center}
\includegraphics[width=\columnwidth]{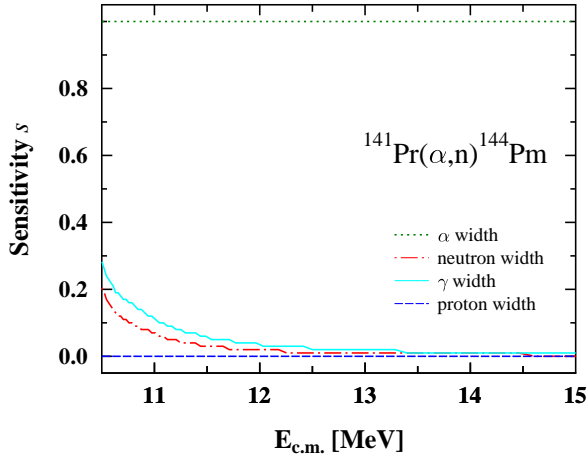}
\end{center}
\caption{\label{fig:Sensitivity_an} Sensitivities $s$ of the $^{141}$Pr($\alpha$,n) laboratory cross sections when varying neutron, proton, $\alpha$, and $\gamma$ widths separately by a factor of 2.}
\end{figure}

As described in the introduction, photodisintegration reactions such as $^{145}$Pm($\gamma$,$\alpha$)$^{141}$Pr play a major role in the $\gamma$ process. A disadvantage common to all direct photodisintegration experiments is the fact that these measurements, in general, can only account for transitions stemming from the ground state of the target nucleus whereas the high temperatures in an explosive astrophysical environment lead to a significant thermal population of excited levels and reactions proceed on nuclei in excited states as well. The modification of the astrophysical reaction rates due to the thermal population is strong, especially at temperatures typical for explosive nucleosynthesis. Stellar photodisintegration rates can differ by several orders of magnitude from laboratory ground-state rates \cite{Raus11}. It can be shown, however, that the stellar effects are smaller in reactions with positive reaction $Q$ value as compared to their counterparts with negative $Q$ value \cite{Raus11,wardfowler}. Furthermore, it is a commonly adopted method when using fits of reaction rates in stellar models to only include the reaction direction with positive $Q$ value and to compute its reverse rate by application of detailed balance \cite{Fowl67,hwfz,whfz,NONSMOKER}. This ensures the fit accuracy and numerical stability in the solution of the coupled differential equations of the reaction network.

Therefore, the preferred experimental strategy for astrophysical implementation would always be to measure reactions with positive $Q$ values. In this manner, measurements can study cross sections as close as possible to those required for the stellar reaction rates. There are comparatively few exceptions to this rule but they always enforce the importance of measuring captures instead of photodisintegrations \cite{kisscoul,raucoul}. For example, both $^{141}$Pr($\alpha$,$\gamma$) and $^{144}$Sm($\alpha$,$\gamma$) have negative $Q$ values but the stellar effects in their rates are smaller than in their reverse rates.

Furthermore, it is very important to measure in the astrophysically relevant energy range or at least as close as possible to those energies. This ensures that dependencies of the predictions on nuclear properties are similar to those appearing in the stellar rates because these sensitivities strongly vary with energy.

The above considerations are also essential in the reaction studied in this work. Although the $^{145}$Pm($\gamma$,$\alpha$) reaction would act at the high plasma temperature in the $\gamma$ process, it is more advantageous to experimentally study the inverse reaction $^{141}$Pr($\alpha$,$\gamma$)$^{145}$Pm. The sensitivities of its laboratory cross sections to a change of the different input parameters are shown in Fig.\ \ref{fig:Sensitivity_ag}.

A detailed discussion of the usefulness of the sensitivity factor can be found in Ref. \cite{Raus11}.
The sensitivity factor $s$ describes the change in the cross section when one of the
nuclear physics input parameters is changed by a factor of $f = \Gamma^{\prime} /\Gamma$, where $\Gamma$ is the width (or transmission coefficient) before a variation and $\Gamma^{\prime}$ the modified width. A sensitivity $s = 1$ means that the cross section is changed by the same factor as the input parameter,
while $s = 0$ signifies that the cross section is not changed at all; that is, the input parameter has no influence on the predicted cross section. Thus, if a width is changed by a factor $f = \Gamma^{\prime} /\Gamma$, the cross section will change by
\begin{equation}
 \frac{\sigma^{\prime}}{\sigma} = s (f - 1) + 1 ~~\mathrm{for} \begin{cases}
\sigma^{\prime} > \sigma ~\mathrm{and}~ \Gamma^{\prime}> \Gamma \\
\sigma^{\prime} < \sigma ~\mathrm{and}~ \Gamma^{\prime} < \Gamma \end{cases}
\end{equation}
and
\begin{equation}
 \frac{\sigma^{\prime}}{\sigma} =  \frac{1} {s (f - 1) + 1} ~~\mathrm{for} \begin{cases}
\sigma^{\prime} < \sigma ~\mathrm{and}~ \Gamma^{\prime}> \Gamma \\
\sigma^{\prime} > \sigma ~\mathrm{and}~ \Gamma^{\prime} < \Gamma \end{cases} \quad.
\end{equation}

In Fig.\ \ref{fig:Sensitivity_ag} the neutron, proton, $\alpha$, and $\gamma$ widths were each varied independently by a factor of $f = 2$.
The energy window relevant for the calculation of the reaction rate is located between 7.2 MeV and 10.3 MeV for $T = 3$ GK \cite{Raus10}.
In this energy region the cross section prediction is almost exclusively sensitive to the variation in the $\alpha$ width. This width, in turn, is determined by the $\alpha$+nucleus optical-model potential.
At energies measurable with the activation method within a reasonable time (well above 10 MeV), the cross-section prediction is additionally sensitive to the $\gamma$ and neutron widths. Thus, it is not easily possible to extrapolate the experimental data towards the astrophysically relevant region without further assumptions. Furthermore, it would be difficult or impossible to disentangle the impact of the different sensitivities if a discrepancy between experiment and predicted cross sections were found.

For this reason, we chose to perform an experiment using the reaction $^{141}$Pr($\alpha$,n)$^{144}$Pm instead. Because of its larger cross section, a shorter half-life of the reaction product $^{144}$Pm, and stronger $\gamma$ intensities in $^{144}$Nd [compared to the ($\alpha$,$\gamma$) reaction and its reaction and decay products], it is possible to measure this reaction down to lower energies. The sensitivity of the ($\alpha$,n) cross section with variation factors $f=2$ is illustrated in Fig.\ \ref{fig:Sensitivity_an}. Except close to the ($\alpha$,n) threshold, the cross sections are only sensitive to the $\alpha$ width. Thus, a comparison of the measured cross sections with calculated ones allows us to study the $\alpha$+nucleus optical-model potential even at energies where this is not possible by $\alpha$ capture. Unfortunately, the ($\alpha$,n) threshold is located above the upper edge of the astrophysical energy window of the capture reaction and therefore the measurements cannot be extended into this region. Nevertheless, the $\alpha$+nucleus optical potential can be investigated at lower energies than previously available for this target. A previous measurement determined ($\alpha$,n) cross sections with the activation method only at higher energies, between 15 and 45 MeV \cite{Afza05}. In addition, their lowest energy data point at 15.71 MeV carries a very large uncertainty on the energy and cannot constrain the predictions. Within this uncertainties it agrees with our data shown in Sec.\ \ref{sec:results}.

It should be mentioned that $\alpha$+nucleus potentials can also be studied in elastic scattering experiments \cite{Kiss11}. However, these experiments sometimes suffer from ambiguous solutions \cite{Fulo01}.

\section{\label{sec:experiment}Experiment}

\begin{figure}
\begin{center}
\includegraphics[width=\columnwidth]{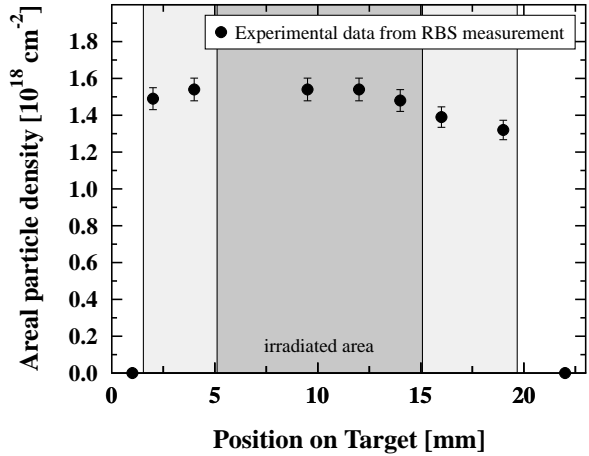}
\end{center}
\caption{\label{fig:homogenity} Homogeneity of the target. The areal particle density of this target was measured at nine positions distributed over the whole target. Two measurements were performed on the Al backing; four on the part onto which Pr was evaporated but which was not irradiated with $\alpha$ particles; three measurements were performed on the irradiated part of the praseodymium. These regions are marked in white, light gray, and dark gray, respectively. The depicted target was irradiated in the activation experiment with $\alpha$ particles of $E_{\alpha}=11$ MeV. The other targets used showed similar homogeneities. }
\end{figure}

\begin{figure}
\begin{center}
\includegraphics[width=\columnwidth]{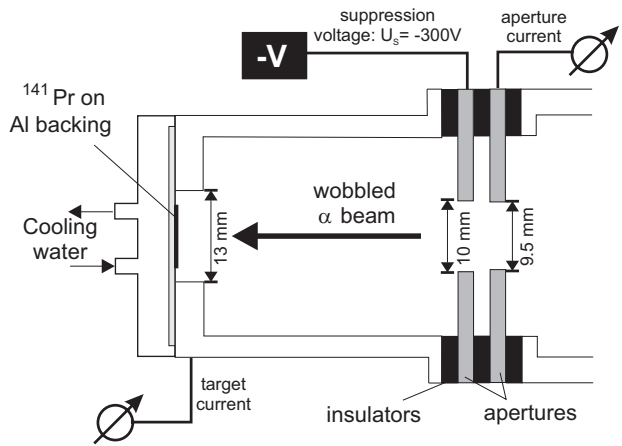}
\end{center}
\caption{\label{fig:setup-PTB} Activation setup at the cyclotron of the PTB Braunschweig. The negatively charged diaphragm ($U_S=-300$ V) suppresses secondary electrons
to ensure that the whole charge is measured on target. To protect the target against decomposition it is water cooled from the outside. }
\end{figure}

All activations were carried out at the cyclotron of the Physikalisch-Technische Bundesanstalt (PTB) in Braunschweig \cite{Bred80}. Thin samples were irradiated for several hours and the $\gamma$ rays emitted after the decay of the radioactive reaction products were detected with a clover-type high purity Germanium (HPGe) detector at the Institut f\"ur Kernphysik in Cologne using the $\gamma \gamma$ coincidence method.

\subsection{\label{target preparation}Preparation and characterization of targets}

Metallic praseodymium was evaporated in vacuum onto 1-mm-thick aluminum disks with a diameter of 35 mm. These backings were thick enough to stop the $\alpha$ beam completely to ensure a reliable charge collection. The Pr samples were characterized by Rutherford back scattering (RBS). The RBS measurements were carried out with singly charged $^4$He ions with an energy of 2 MeV $\pm$ 1 keV at the RUBION Dynamitron-tandem accelerator laboratory at the Ruhr University of Bochum. The targets were mounted on a movable sample holder, which serves as a Faraday cup and were irradiated with a beam current of 14 nA. Secondary electrons were suppressed by an appropriate voltage ($-300$ V) to ensure a reliable charge collection using a calibrated integrator. The silicon detector was mounted at a distance of 35 mm from the center of the sample. Its resolution is 15 keV for 2 MeV $\alpha$ particles. The solid angle of the detector was determined to be $(1.91 \pm 0.07)$ msr. The areal particle density of the different targets is between 1.36 and 1.88 $\cdot 10^{18} \cdot \mathrm{cm}^{-2}$ which corresponds to an areal density between ($318 - 440$) $\mu$g/cm$^2$. To study the homogeneity of the targets, the areal particles densities of several targets were determined at many points distributed over the target area. An example is displayed in Fig.\ \ref{fig:homogenity}. A second RBS measurement after the activation experiment has shown that within the uncertainties, no target material was lost during the activation runs.

The metallic Pr targets oxidize completely within a few days. Because the oxide is not bound on the backing anymore, it is mandatory to avoid oxidation before the activation and to fix the target material at its position if oxidation occurs. Therefore, the targets were stored in vacuum before the irradiations. After the irradiations they were put in a paraffin envelope.

\subsection{\label{experimental setup}Experimental installation at PTB}

Doubly charged He$^{2+}$ ions were extracted from the cyclotron to irradiate the $^{141}$Pr targets in an activation chamber, which is designed as a Faraday cup. The charge deposited on the target was recorded in time steps of 60 seconds by a current integrator for later correction of beam-current fluctuations. The beam-current integration at this setup is well established and an uncertainty of 1\% has been taken into account. Secondary electrons were suppressed by a negatively charged diaphragm ($U_S=-300$ V) at the entrance of the activation chamber.
A homogenous illumination of the target was achieved by wobbling the $\alpha$ beam. For each energy the wobbling was optimated by inserting a quartz window at the target position and checking its homogeneous irradiation.
The beam spot was about ten mm in diameter. The backings were water cooled from the outside. In Fig.\ \ref{fig:setup-PTB} a schematic of the target chamber at PTB is shown. The energy of the $\alpha$-beam was defined within an uncertainty of $\pm 25$ keV by means of the field calibration of two analyzing magnets as well as by a time-of-flight measurement of the particle velocity \cite{Boet02}.

\subsection{\label{irradiation}Irradiation and $\gamma$ counting}

In total ten $^{141}$Pr samples were irradiated at eight different $\alpha$ energies between 11 MeV and 15 MeV to produce the unstable reaction product $^{144}$Pm.
In order to test the thermodynamical stability of the targets, two targets were irradiated with different beam currents between 0.08 and 3.5 $\mu$A at $\alpha$ energies of 12.6 MeV and 15 MeV. These measurements yielded current-independent cross sections within the uncertainties, excluding decomposition of target material. The Q value of the $^{141}$Pr($\alpha$,n)$^{144}$Pm reaction is $Q = (-10 246.19 \pm 2.70)$ keV and the reaction threshold amount to $(10 537.23 \pm 2.78)$ keV \cite{NNDC}. The duration of the activation runs was varied between one hour and 17 hours. The average beam current was 3 $\mu$A.

The produced activity of all targets was measured off beam at the Institut f\"ur Kernphysik in Cologne using a clover-type HPGe detector with a relative efficiency of 120\% at $E_{\gamma}=1332.5$ keV, compared to a $3\times 3 $ inch cylindrical NaI detector. The activity was determined by performing spectroscopy of the $\gamma$ rays emitted after the electron capture of $^{144}$Pm with a low-background counting setup. The clover-type HPGe detector is composed of four crystals, which provides the possibility of measuring coincidences between the crystals. Since the ($\alpha$,n) cross section at $E_\alpha$= 11 MeV is in the range of 10 $\mu$b only and the reaction product $^{144}$Pm decays with a long half-life of $T_{1/2} = (363 \pm 14)$ days \cite{NNDC}, the count rate in the singles spectra is below or close to the sensitivity limit. Therefore, the requirement of coincidences is mandatory to enhance the peak-to-background ratio. The duration for the counting lasted between one and 21 days. In the following, this detector is referred to as the Cologne clover.

In order to exclude systematic errors in the $\gamma \gamma$ coincidence method, singles spectra of five targets were recorded additionally at PTB with a coaxial HPGe detector with a relative efficiency of 70\% (PTB 70\% detector). The Cologne clover as well as the PTB 70\% detector were used in a close geometry between the target and the detector end cap. Moreover, the decay of the reaction product involves three cascading $\gamma$ transitions. Therefore, coincidence summing effects, which occur when two or more $\gamma$ rays are recorded in one crystal within the resolving time of this crystal, have to be taken into account when using a close geometry \cite{Debe88}. Therefore, the efficiencies of these detectors were corrected for coincidence summing effects (see Sec.\ \ref{sec:Efficiencies}).

To verify this correction, a further measurement was performed at the Institut f\"ur Kernphysik in Cologne with a large distance between end cap and target by using a coaxial HPGe detector with a relative efficiency of 55\%. Due to this large distance, summing effects play a marginal role. This detector is named Cologne detector in the following.

Furthermore, it was possible to perform the spectroscopy of the target which was irradiated with an energy of $E_\alpha = 11.4$ MeV at the underground laboratory for dosimetry and spectrometry (UDO) at the Asse salt mine near Braunschweig. PTB operates various low-background $\gamma$-ray spectrometry systems at this underground laboratory at a depth of 1200 m water equivalent. A coaxial ultralow-background HPGe detector with a relative efficiency of 90\% (ULB) was used for this measurement. Detailed information about the ULB detector and the UDO laboratory can be found in Ref. \cite{Neum09}. This measurement was analyzed completely independently from the other measurements mentioned above. Two of the targets were measured in addition at the PTB with a coaxial HPGe detector with a relative efficiency of 50\% (PTB 50\% detector) and were analyzed with the same routine as the one which was counted at the ULB detector.

Within the uncertainties the measurements with different detectors gave consistent results.

\section{\label{sec:data analysis}Data Analysis}

\begin{figure}
\begin{center}
\includegraphics[width=\columnwidth]{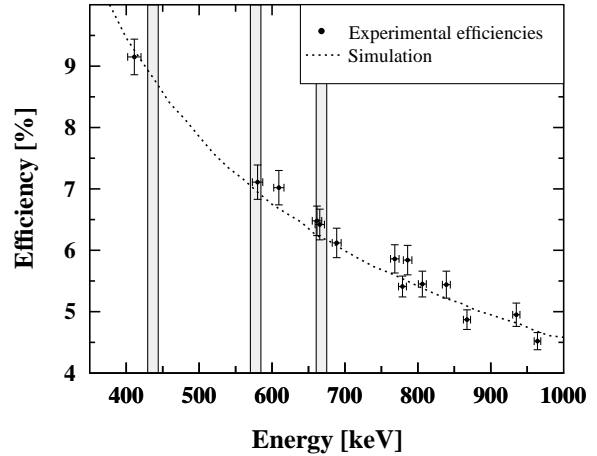}
\end{center}
\caption{\label{fig:efficiencies} Comparison between the experimentally determined efficiencies without summing effects and the simulation for the Cologne clover. It can be seen that the {\sc Geant4} simulation describes the experimental efficiencies very well if summing effects are negligible. Only the energy region of interest is shown here and the energy region where the transitions in $^{144}$Nd occur is marked by the shaded area.}
\end{figure}

\begin{figure}
\begin{center}
\includegraphics[width=\columnwidth]{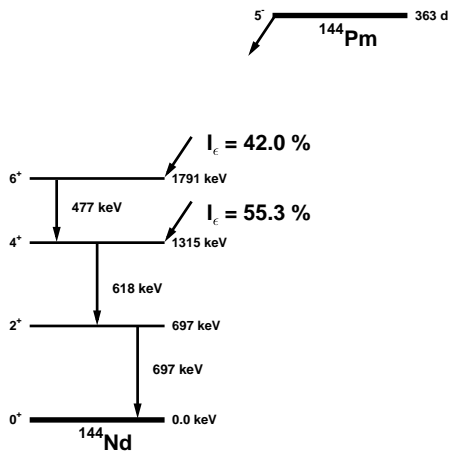}
\end{center}
\caption{\label{fig:level} Simplified decay scheme of $^{144}$Pm. Only transitions which were used for the data analysis are shown. Spins, parities, energies and population probabilities $I_{\epsilon}$ are taken from \cite{NNDC}.}
\end{figure}

The reaction product $^{144}$Pm decays via electron capture with a half-life of $T_{1/2}= (363 \pm 14)$ days to $^{144}$Nd \cite{NNDC}, see Fig. \ref{fig:level}. Subsequent to the decay, the reaction product is in an excited state, which deexcites via the emission of $\gamma$ rays. The three strongest $\gamma$ transitions used for the analysis proceed in a cascade; their properties are summarized in Table \ref{tab:decay}.

\begin{table}
\caption{\label{tab:decay} Decay data of the reaction product $^{144}$Pm. Only transitions which were used for the data analysis are listed. The decay parameters are taken from \cite{NNDC}.}
\begin{ruledtabular}
\begin{tabular}{ccc}
\textrm{$E_\gamma$ / keV}&
\textrm{$I_\gamma$}&
\textrm{Mult.}\\
\colrule
 & & \\
476.78 $\pm$ 0.03  & 0.4378 $\pm$ 0.0199  & \textrm{E2}\\
618.01 $\pm$ 0.03 & 0.9850 $\pm$ 0.0298  & \textrm{E2}\\
696.49 $\pm$ 0.03 & 0.9949 $\pm$ 0.0002 & \textrm{E2}\\
\end{tabular}
\end{ruledtabular}

\end{table}

The counts in the full-energy peak of a certain transition $Y(E_\gamma)$ are connected with the total number of decays of the reaction product $\Delta N$ during the counting time by the following expression:
\begin{equation}
 Y(E_\gamma) = I_\gamma(E_\gamma) \epsilon(E_\gamma) \frac{t_\mathrm{LIVE}}{t_\mathrm{REAL}} \Delta N \quad.
\end{equation}
The full-energy efficiency is denoted as $\epsilon(E_\gamma)$ and the absolute $\gamma$ intensity per decay of the mother nuclide $^{144}$Pm is indicated by $I_\gamma(E_\gamma)$. The correction $\frac{t_\mathrm{LIVE}}{t_\mathrm{REAL}}$ takes the dead time of the measurement into account, which was smaller than 1 \% for all targets.

From the decay law another relation for the total number of decays of $^{144}$Pm during the counting time can be derived:
\begin{equation}
 \Delta N  = N_\mathrm{act}  \left[ 1- e^{-\lambda  \Delta t_\mathrm{meas}} \right]  e^{-\lambda  \Delta t_\mathrm{wait}} \quad.
\end{equation}
The total number of $^{144}$Pm nuclei at the end of the activation is $N_\mathrm{act}$. The counting time is denoted by $\Delta t_\mathrm{meas}$, whereas the time between the end of the activation and the beginning of the counting is denoted as $\Delta t_\mathrm{wait}$. The quantity $\lambda$ is the decay constant.

During the activation $N_\mathrm{prod}$ $^{144}$Pm nuclei are produced in total but, during the activation, a part of these nuclei decay so that, as mentioned before, $N_\mathrm{act} $ reaction products are left at the end of the activation:
\begin{equation}
  N_\mathrm{act} = f_\mathrm{act}  N_\mathrm{prod} \quad.
\end{equation}
For a constant beam current during the activation the factor $ f_\mathrm{act}$ is obtained by
\begin{equation}
 f_\mathrm{act} = \frac{\left( 1 - e^{-\lambda  \Delta t_\mathrm{act}} \right)}{\Delta t_\mathrm{act} \lambda} \quad.
\end{equation}
The activation duration is denoted as $\Delta t_\mathrm{act}$.
The beam current was recorded in time intervals of 60 s and an appropriate correction for small fluctuations was applied.

\subsection{\label{sec:Efficiencies}Determination of detector efficiencies}

For the calculation of the reaction cross section the absolute full-energy efficiencies of the detectors have to be known. In total, five HPGe detectors were used for the spectroscopy as described in Sec.\ \ref{irradiation}. The efficiencies of all detectors were determined and the procedure is presented in the following.

Because of the low activity of the targets, the distance between the target and the end cap of the Cologne clover was only 5.9 mm. As mentioned before, the decay of $^{144}$Pm involves three cascading $\gamma$ transitions, so that coincidence summing effects have to be taken into account. This holds for any multiline calibration source with cascading $\gamma$ transitions.

The absolute detector efficiency was measured in far geometry (10 cm distance between end cap of the detector and calibration source), where coincidence summing is negligible. In total, six calibration sources ($^{60}$Co,$^{137}$Cs, $^{152}$Eu, $^{133}$Ba, $^{57}$Co, and $^{226}$Ra) were used.

In a second step, the efficiency was measured for all six calibration sources in close geometry as well. For the $\gamma$ transition following the decay of $^{137}$Cs a conversion factor was calculated between the efficiency in far and in close geometries. This is a pure geometric factor, because no coincidence summation occurs.  Simulations with {\sc Geant4} \cite{geant4} confirm that the assumption of an energy-independent conversion factor is valid in an energy range between 200 keV and 2000 keV. Thus, the experimental efficiency curve determined in far geometry is shifted by this factor to yield the efficiencies in the close geometry without any summing effects.

Monte Carlo simulations with {\sc Geant4} were performed for this close geometry and were compared to the experimental efficiencies without summing effects. The simulation describes the experimental efficiencies very well. This is shown in Fig.\ \ref{fig:efficiencies}.

In a third step, we simulated the extended geometry of our targets but, within the uncertainties, the simulation delivers the same results for a point source and our target geometry.

In the last step, the $\gamma$ cascade of $^{144}$Nd was implemented into the simulation to take summing effects into account. The simulation does not include $\gamma \gamma$ angular correlation between the three $\gamma$ rays. Calculations have shown that the effect of this $\gamma \gamma$ angular correlation on the efficiency is smaller than 1\%.

The counting with the PTB 70\% detector was performed in the close geometry as well. Therefore, the same procedure as for the Cologne clover was used for the determination of the absolute full-energy efficiency of the PTB 70\% detector. In this case only three calibration sources ($^{60}$Co, $^{152}$Eu, and $^{137}$Cs) were used.

The efficiencies of the ULB detector and the PTB 50\% detector were simulated with Gespecor \cite{Sima00,Sima01}. This program uses recent data from the data center DDEF \cite{DDEF}. Summing corrections were also calculated with special routines of this code. For the rest analysis of the ULB spectra the program ``Genie 2000 Gamma Analysis Software'' by CANBERRA was applied, whereas the data of the PTB 50\% detector were analyzed by using the program Interwinner by Ortec.

Finally, a large distance of 9.5 cm between end cap and target was used with the Cologne detector. Therefore, summing effects play a marginal role and simulations with {\sc Geant4} were unnecessary. An interpolation between the measured efficiencies using two calibration sources ($^{137}$Cs and $^{152}$Eu) was sufficient to obtain the full-energy efficiencies at energies of the $^{144}$Pm emission lines.

\subsection{$\gamma \gamma$ coincidence method}

\begin{figure}
\begin{center}
\includegraphics[width=\columnwidth]{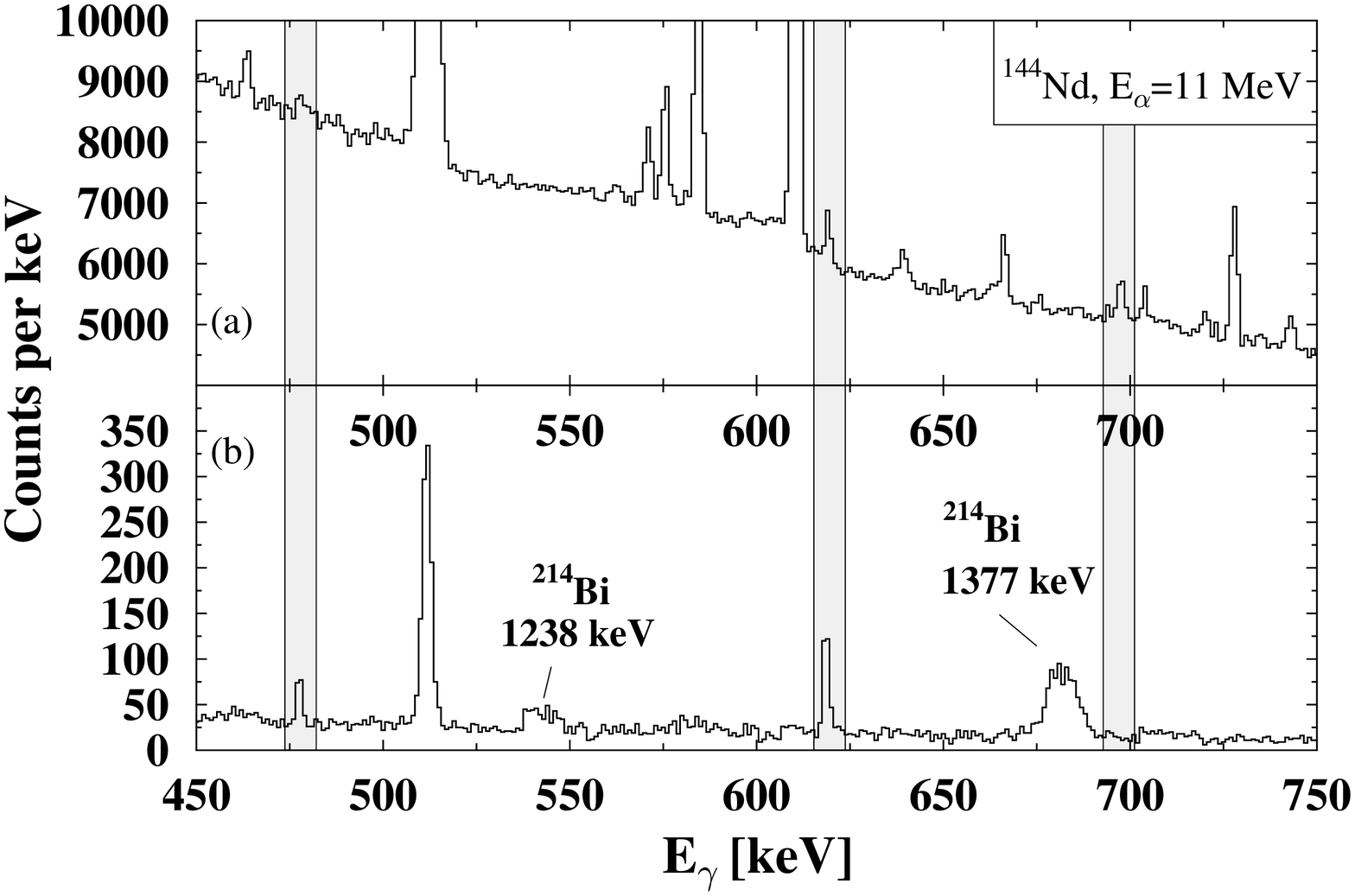}
\end{center}
\caption{\label{fig:coin} Sensitivity of the $\gamma \gamma$ coincidence method. In the upper panel (a) the singles spectrum for a target irradiated with 11$~$MeV $\alpha$ particles is shown, whereas the lower panel (b) shows the corresponding coincidence spectrum for a gate triggered by 696 keV photons.}
\end{figure}

As mentioned above, the activity of the targets irradiated with $\alpha$ particles at 11 MeV is below or of the order of the background activity. Therefore, even the strongest transitions in $^{144}$Nd are only weakly visible in the singles spectra of the Cologne clover. A $\gamma \gamma$ coincidence method was applied to suppress the background.

All coincidences between two crystals in the list mode data were sorted into a coincidence matrix. This symmetric matrix contains all coincidence events of any crystal pairs. An example for the analyzing power of the coincidence technique is depicted in Fig.\ \ref{fig:coin}.

In Fig. \ref{fig:coin}(a) the singles spectrum of the target irradiated with 11 MeV $\alpha$ particles is shown, whereas Fig. \ref{fig:coin}(b) shows the corresponding coincidence spectrum after applying a gate triggered by 696 keV photons. Both spectra show the relevant energy range, where all three transitions in $^{144}$Nd are located. The energies of the transitions are marked in gray. In the singles spectrum the $\gamma$ lines are hardly visible and superimposed on a huge background. In the coincidence spectrum the transitions are clearly visible and the peak-to-background ratio is improved dramatically compared to the singles spectrum. The broad lines in the coincidence spectrum result from Compton scattering of strong background lines from $^{214}$Bi, when 696 keV of the total energy is deposited in one of the crystals and an energy of $E_\gamma=E_\mathrm{total} - 696$ keV is deposited in another one \cite{Daba04}.

To determine the total cross section, the absolute full-energy efficiency for the coincidence spectrum is needed as well. Five targets were used, two of them were bombarded at 15 MeV, the others were bombarded at 14.4 MeV, 13.8 MeV, and 13.2 MeV, respectively. Conversion factors $\eta(E_\mathrm{gate}, E)$ between the singles spectrum and the coincidence spectrum were then calculated by
\begin{equation}
 \eta(E_\mathrm{gate}, E) = \frac{N_\mathrm{single}(E)}{N_\mathrm{coin}(E_\mathrm{gate},E)}   \quad.
\end{equation}
The yield in the singles spectrum at an energy $E$ is denoted $N_\mathrm{single}(E)$ and the yield in the coincidence spectrum at an energy $E$ gated by photons of an energy $E_\mathrm{gate}$ is $N_\mathrm{coin}(E_\mathrm{gate},E) $. For each of the three transitions two conversion factors were calculated depending on which coincidence pair is considered. In total, six conversion factors were obtained by averaging $\eta(E_\mathrm{gate}, E)$ over the five targets. These conversion factors, which are independent of the target activity, were used to calculate the corresponding yields in the singles spectrum $N_\mathrm{single}(E)$ for all targets. These yields are used to derive the cross sections for the Cologne clover as described in Sec.\ \ref{sec:data analysis}.

\section{\label{sec:results}Results and Discussion}

\begin{table*}
\caption{\label{tab:results} Summary of the experimental cross sections listed for each energy $E_\alpha$ of the $\alpha$ particles together with the detectors used. The areal particle density $m$ of the targets is also indicated.}
\begin{ruledtabular}
\begin{tabular}{cccc}
 \textrm{$E_\alpha$}&
\textrm{$m$}&
\textrm{Detectors Used} &
$\sigma$ \\
 keV & $\mathrm{cm}^{-2}$& &  mb \\
\colrule
                      &                                         &                               &               \\
11047 $\pm$ 28  & $(1.32 \pm 0.05) \cdot 10^{18} $ & \textrm{Cologne clover} & $0.008 \pm 0.001$\\ \hline
                      &                                         &                               &               \\
11305 $\pm$ 28  & $(1.42 \pm 0.06) \cdot 10^{18} $ & \textrm{Cologne clover} & $0.015 \pm 0.001$\\
                  &                                             &\textrm{ULB}   &       $0.015 \pm 0.001$               \\
                  &                                             &\textrm{PTB 50$\%$ detector}       &       $0.014 \pm 0.001$               \\ \hline
                      &                                         &                               &               \\
11899 $\pm$ 29  & $(1.71 \pm 0.07) \cdot 10^{18} $ &\textrm{Cologne clover}  & $0.07 \pm 0.01$\\ \hline
                      &                                         &                               &               \\
12528 $\pm$ 29  & $(1.80 \pm 0.07) \cdot 10^{18} $ & \textrm{Cologne clover} & $0.31 \pm 0.02$\\ \hline
                      &                                         &                               &               \\
12549 $\pm$ 28  & $(1.36 \pm 0.05) \cdot 10^{18} $ & \textrm{Cologne clover} & $0.35 \pm 0.02$\\
                &                                               &\textrm{PTB 70$\%$ detector}          & $0.35 \pm 0.03$\\ \hline
                      &                                         &                               &               \\
13098 $\pm$ 29  & $(1.88 \pm 0.08) \cdot 10^{18} $ & \textrm{Cologne clover} & $1.20 \pm 0.07$\\
                  &                                             &\textrm{PTB 70$\%$ detector}          & $1.20 \pm 0.10$\\ \hline
                      &                                         &                               &               \\
13736 $\pm$ 28  & $(1.56 \pm 0.06) \cdot 10^{18} $ & \textrm{Cologne clover} & $3.58 \pm 0.24$\\
                &                                               & \textrm{PTB 70$\%$ detector}         & $3.56 \pm 0.31$\\ \hline
                      &                                         &                               &               \\
14426 $\pm$ 29  & $(1.61 \pm 0.06) \cdot 10^{18} $ & \textrm{Cologne clover} & $11.02 \pm 0.68$\\
                &                                               & \textrm{PTB 70$\%$ detector}         &$10.95 \pm 0.91$\\
                  &                                             &\textrm{PTB 50$\%$ detector}       &       $9.98 \pm 0.60$       \\
                &                                               & \textrm{Cologne detector}     & $11.40 \pm 0.88$\\ \hline
                      &                                         &                               &               \\
14956 $\pm$ 28  & $(1.72 \pm 0.07) \cdot 10^{18} $ & \textrm{Cologne clover} & $23.30 \pm 1.37$\\
                &                                               & \textrm{PTB 70$\%$ detector}         & $23.69 \pm 1.96$\\ \hline
                      &                                         &                               &               \\
14955 $\pm$ 28  & $(1.65 \pm 0.07) \cdot 10^{18} $ & \textrm{Cologne clover} & $26.29 \pm 1.68$\\
                &                                               & \textrm{PTB 70$\%$ detector}         &$24.07 \pm 2.00$\\
\end{tabular}
\end{ruledtabular}
\end{table*}


\begin{table}
\caption{\label{tab:sfactors}Astrophysical $S$ factors as a function of center-of-mass energy, calculated from the cross sections averaged over all detectors and evaluated at the lower and upper edges of the energy uncertainties $E_\mathrm{S}=E_\mathrm{c.m.} \pm \Delta E$ from the averaged uncertainties. The uncertainties $\Delta S$ on the $S$ factors at each energy $E_\mathrm{c.m.} \pm \Delta E$ are derived including the averaged uncertainties of the cross section. The uncertainties are quoted for a one $\sigma$ confidence inteval (coverage factor k=1).}
\begin{ruledtabular}
\begin{tabular}{ccccc}
\multicolumn{1}{c}{$E_\mathrm{c.m.}$}&\multicolumn{1}{c}{$\Delta E$}&\multicolumn{1}{c}{$E_\mathrm{S}$}&\multicolumn{1}{c}{$S$ factor}&\multicolumn{1}{c}{$\Delta S$} \\
\multicolumn{1}{c}{MeV}&\multicolumn{1}{c}{MeV}&\multicolumn{1}{c}{MeV}&\multicolumn{1}{c}{10$^{26}$ MeVb}&\multicolumn{1}{c}{10$^{26}$ MeVb} \\
\hline
10.742 & 0.02818 && \\
&&  10.714  &  3.0349 & 0.37010   \\
&&  10.770  &  2.5375 & 0.30945  \\
10.993 & 0.02828 &&  \\
&&  10.965  &  2.4108 & 0.21916 \\
&&  11.021  &  2.0282 & 0.18438 \\
11.570 & 0.00289 &&  \\
&&  11.541  &  3.1276 & 0.21116 \\
&&  11.599  &  2.6545 & 0.17922 \\
12.202 & 0.02796 && \\
&&  12.174  &  1.8769 & 0.09157 \\
&&  12.230  &  1.6213 & 0.07910 \\
12.182 & 0.02916 && \\
&&  12.153  &  1.7686 & 0.10828 \\
&&  12.211  &  1.5176 & 0.09291 \\
12.736 & 0.02890 && \\
&&  12.707  &  1.6745 & 0.10099 \\
&&  12.765  &  1.4531 & 0.08764 \\
13.357 & 0.02841 && \\
&&  13.329  &  1.1414 & 0.07502 \\
&&  13.385  &  1.0025 & 0.06590 \\
14.028 & 0.02864 && \\
&&  13.999  &  0.78233 & 0.04914 \\
&&  14.057  &  0.6929 & 0.04352 \\
14.544 & 0.02803 && \\
&&  14.516  &  0.5853 & 0.03425 \\
&&  14.572  &  0.52304 & 0.03061 \\
14.542 & 0.02841 && \\
&&  14.514  &  0.63701 & 0.04211 \\
&&  14.570  &  0.56837 & 0.03757
\end{tabular}
\end{ruledtabular}
\end{table}

\begin{figure}
\begin{center}
\includegraphics[width=\columnwidth]{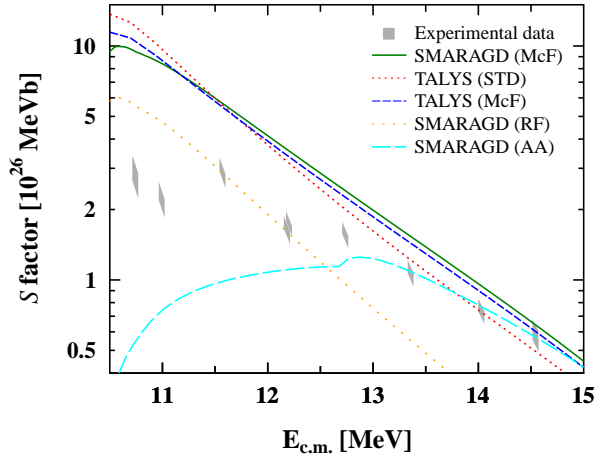}
\end{center}
\caption{\label{fig:WQS} Astrophysical $S$ factors for reaction $^{141}$Pr($\alpha$,n)$^{144}$Pm as function of c.m.\ energy. The experimental $S$ factors were computed from averaged cross section values from all detectors. These are compared to calculations with the codes SMARAGD \cite{Raus11,SMARAGD} and TALYS \cite{TALYS, TALYS11} using $\alpha$+nucleus optical-model potentials from \cite{McFa66} (McF), \cite{avri10} (AA), \cite{frohdip,raufroh} (RF), and the standard potential \cite{Wata58, Madl88} used in TALYS (STD).}
\end{figure}

\begin{figure}
\begin{center}
\includegraphics[angle=0,width=\columnwidth]{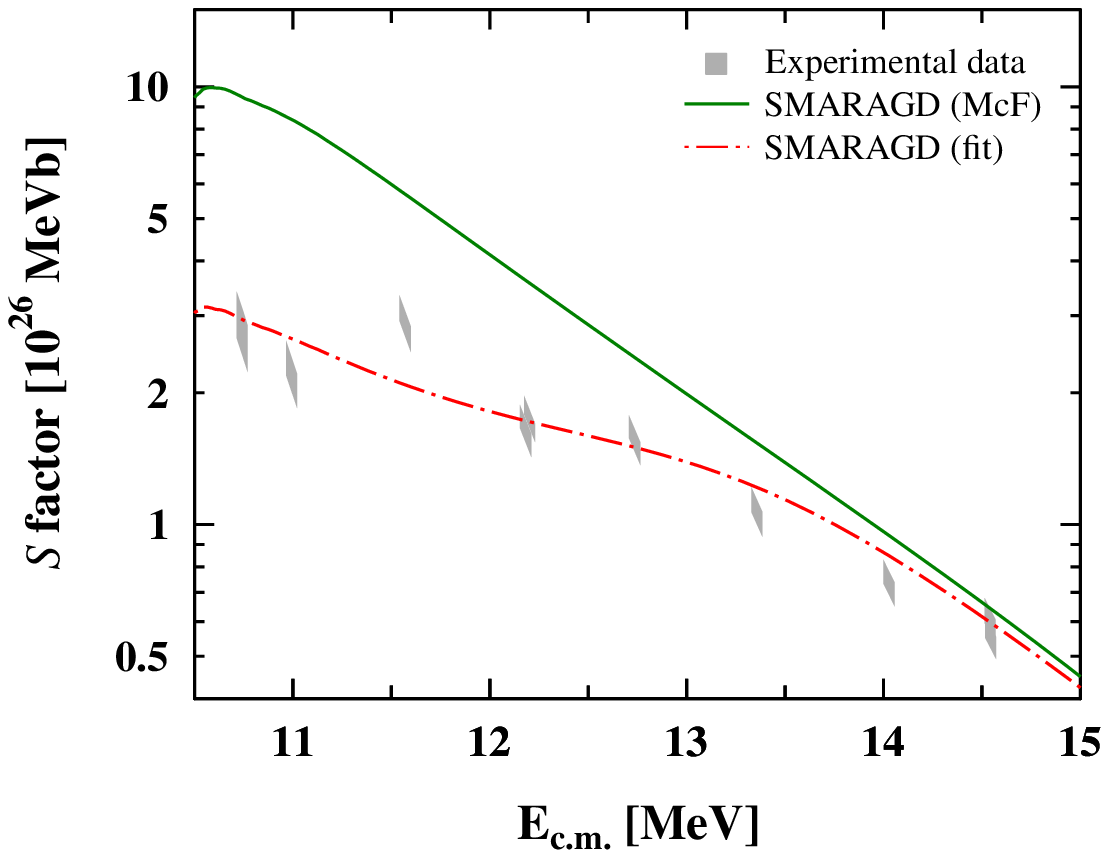}
\end{center}
\caption{\label{fig:fit} Astrophysical $S$ factors for reaction $^{141}$Pr($\alpha$,n)$^{144}$Pm as function of c.m.\ energy. The experimental $S$ factors were computed from averaged cross section values from all detectors. These are compared to calculations with the codes SMARAGD \cite{Raus11,SMARAGD} using the $\alpha$+nucleus optical-model potentials from \cite{McFa66} (McF) and a potential with an energy-dependent depth of the imaginary part (fit).}
\end{figure}

The experimentally determined cross sections obtained with each detector used in this experiment are shown in Table \ref{tab:results}. The values for the different detectors are in excellent agreement. This proves that the $\gamma \gamma$ coincidence method can be used to determine absolute values for cross sections in activation experiments.

The energy $E_\alpha$ of the $\alpha$ particles was obtained by correcting the adjusted primary energy $E_0$ of the $\alpha$-particle beam with the energy loss $\Delta E$ inside the target material
\begin{equation}
 E_\alpha = \frac{E_0 + (E_0 -\Delta E)}{2} \quad .
\end{equation}
This energy determination, which is appropriate if the energy loss and the cross sections change only slightly over the target thickness, is independent from a specific cross section prediction.  In this experiment the uncertainties of the cross section are larger than the changes in the cross section over the target thickness. The widths of the energies is determined by the straggling in the target and the uncertainty of the beam energy of the accelerator. They were added according to Gaussian error propagation. As mentioned in Sec. \ref{experimental setup}, the energy uncertainty of the cyclotron at the PTB is $\pm$ 25 keV. The energy loss of the $\alpha$ particles in the target itself is obtained by a {\sc Geant4} simulation and varied between 62 keV and 85 keV. The simulation yields the distribution of the $\alpha$-particle beam after traveling through the target material. The maximum of this distribution is $E_0-\Delta E$ and the halfwidth of this distribution at $1/e$ of the maximum is the straggling in the target which varied between 13 keV and 15 keV.

Table \ref{tab:sfactors} gives the astrophysical $S$ factors obtained with weighted averages of the cross sections from the different detectors shown in Table \ref{tab:results}. Note that the energies in Table \ref{tab:sfactors} are given as center-of-mass (c.m.) energies (as in Figs.\ \ref{fig:Sensitivity_ag}, \ref{fig:Sensitivity_an}, \ref{fig:WQS}, and \ref{fig:fit}) to facilitate comparison with theoretical calculations. Since the energy enters the calculation of the $S$ factor, the inclusion of the uncertainties on cross sections and energies is not straightforward. A pair of $S$ factors is given for each c.m.\ energy $E_\mathrm{c.m.}$ in Table \ref{tab:sfactors}, evaluated at the lower and upper limit of the energy range defined by the uncertainties $E_\mathrm{c.m.} \pm \Delta E$. The error bars on $\alpha$ energy and cross section translate into an error trapezoid for the $S$ factor, with its four corners given by the upper and lower limit of the $S$ factor pair at each c.m.\ energy. This error trapezoid is also shown in Figs.\ \ref{fig:WQS} and \ref{fig:fit}.

As discussed in Sec.\ \ref{sec:sensitivities}, the cross section of the ($\alpha$,n) reaction is almost exclusively sensitive to a change in the $\alpha$ width (or equivalently the total $\alpha$ Hauser-Feshbach transmission coefficient). Entering the calculation of the $\alpha$ width are the transmission coefficients for the individual $\alpha$ emissions to the ground and excited states of $^{141}$Pr which are computed by solving radial Schr\"odinger equations using an $\alpha$+nucleus potential and the appropriate quantum numbers of the involved states \cite{Raus11}. Therefore, the excited states in $^{141}$Pr also have to be known or a nuclear level density where the discrete states are unknown. In the laboratory cross sections and $S$ factors discussed here, the $\alpha$ widths are dominated by transitions to low-lying excited states because transitions to levels at higher excitation energies have lower relative $\alpha$ energy and are strongly suppressed by the Coulomb barrier. The properties of the relevant states are mostly known and therefore the only uncertainty arises from the $\alpha$+nucleus optical-model potential (OMP).

Figure \ref{fig:WQS} shows a comparison between the experimental $S$ factors from Table \ref{tab:sfactors} and calculations performed with the statistical model codes SMARAGD \cite{Raus11,SMARAGD} and TALYS \cite{TALYS,TALYS11} using different $\alpha$+nucleus OMPs.
Both codes were used with their default settings; only the $\alpha$+nucleus OMP was varied. This figure can be directly compared with Fig.\ \ref{fig:Sensitivity_an} to judge the impact of the OMP. Only below 11 MeV -- close to the ($\alpha$,n) threshold -- may variations in the $\gamma$ and neutron widths also influence the prediction to a small extent, as well as threshold effects like width fluctuations.
Interestingly, none of the used OMPs can reproduce the energy dependence of the data, even above 11 MeV.

The most widely used astrophysical reaction rates include the OMP by \cite{McFa66} (denoted by 'McF' in Fig. \ref{fig:WQS}). This potential was obtained by fitting a four-parameter Woods-Saxon potential on measured $\alpha$-particle elastic scattering data at an energy of $E_{\alpha}=24.7$ MeV on nuclei between oxygen and uranium. It was successful in reproducing a wide range of scattering and reaction data at higher energies but was found to systematically overpredict reaction data close to the astrophysically relevant energy region (see, e.g., \cite{tm169,yalcin,gyurky,151eu,rapp02,rapp08,somo98} and references therein). Here, we find similar overpredictions from a factor of two in the region $11.5\leq E_\mathrm{c.m.}^{\alpha}\leq 12.5$ MeV to a factor of four below 11 MeV. Only the data at the upper end of the measured energy range is reproduced satisfactorily with this potential. This is interesting as it shows the transition from an energy region where the potential is appropriate to the region where it is not.

By default, the TALYS code uses a simplification of the folding approach by \cite{Wata58, Madl88} for the $\alpha$ OMP (denoted by 'STD' in Fig.\ \ref{fig:WQS}). The resulting energy dependence of the $S$ factors is slightly steeper than the one with \cite{McFa66}. It shows comparable discrepancy with the data at the lowest energies but reproduces the data at higher energy slightly better. It seems, however, as if the disagreement worsens again when continuing to even higher energies above the measured range.

Motivated by the failing of the OMP by \cite{McFa66}, a modified potential was derived in Refs. \cite{frohdip,raufroh} by simultaneously fitting low-energy reaction data in the $\mathrm{A} \simeq 140$ mass region; more specifically, reaction data of $^{143}$Nd(n,$\alpha$)$^{140}$Ce \cite{raufroh}, $^{147}$Sm(n,$\alpha$)$^{144}$Nd \cite{Gled01}, and $^{144}$Sm($\alpha$,$\gamma$)$^{148}$Gd \cite{somo98}. This OMP has the same imaginary part as \cite{McFa66} but a shallower real part, resulting in generally lower cross sections. It does not describe scattering data but that was never intended. The original notion was that the potential should only be applied at astrophysically low energies (close to the Coulomb barrier) although an explicit energy dependence of its parameters was not introduced due to lack of data. The comparison in Fig.\ \ref{fig:WQS} shows that indeed the higher-energy region is not reproduced well whereas the calculated $S$ factors are closer to the experimental ones than those obtained with the other OMPs at lower energy. Interestingly, the energy dependence of the $S$ factors calculated with this potential is very similar to the one obtained with the standard TALYS potential but the values are shifted in magnitude.

The idea of \cite{frohdip,raufroh} to include reaction data into fits of the $\alpha$ OMP was recently picked up again by \cite{avri10}. They fit mass- and energy-dependent potential parameters to ($\alpha$,$\gamma$), ($\alpha$,n), and ($\alpha$,xn) reaction cross sections closely around the Coulomb barrier for targets in the range $121\leq A\leq 197$ and also attempted to reproduce scattering data available at higher energy. This OMP includes a volume and a surface term in the imaginary part (both mass- and energy-dependent) and has 48 parameters in total, which is to be compared to the 7 parameters of the potentials by \cite{McFa66,frohdip,raufroh}. It has to be noted that the $^{141}$Pr($\alpha$,n) reaction data at higher energy by \cite{Afza05} were included in the fit. The general difficulty of extrapolating potential parameters to low energies is illustrated by the comparison of the $S$ factors obtained with this OMP to our data in Fig.\ \ref{fig:WQS} (the OMP is marked by 'AA' in this figure). The three data points at the highest measured energies are reproduced perfectly but the $S$ factors drop much too rapidly with decreasing energy below about 13 MeV.

\begin{table}
\caption{\label{tab:omppars}Parameters for the energy-dependent $\alpha$+nucleus optical-model potential derived from the data in this work.}
\begin{ruledtabular}
\begin{tabular}{rrlrrl}
\multicolumn{3}{c}{Real part}&\multicolumn{3}{c}{Volume imaginary part}\\
\hline
$r_\mathrm{C}=$&1.20& fm & &&\\
$V=$&185.00& MeV & $W=$&$\frac{25}{\left\{1+\exp\left[\left(0.9E_\mathrm{C}-E_\mathrm{c.m.} \right)/2\right]\right\}}$& MeV \\
$r_\mathrm{R}=$&1.40& fm & $r_\mathrm{V}=$&1.40& fm \\
$a_\mathrm{R}=$&0.52& fm & $a_\mathrm{V}=$&0.52& fm
\end{tabular}
\end{ruledtabular}
\end{table}

As previously mentioned, the data in this work seem to cover regions where the standard potentials work acceptably well and such where they fail. It can be attempted to bridge the two regimes by introducing a simple energy-dependence connecting the potentials by \cite{McFa66} and \cite{frohdip,raufroh}, following up on the original idea. The example of the OMP by \cite{avri10}, however, shows that one has to be cautious and that such extrapolations may not be generally valid. Therefore, we tried to limit the number of parameters as much as possible. They are listed in Table \ref{tab:omppars}. It was chosen to keep the depth of the OMP real part $V$ and the geometry of real and imaginary part of \cite{McFa66}. Only the depth of the volume imaginary part $W$ was made energy dependent. It has to approach the value of \cite{McFa66} at high energy but has to be shallower at energies below the Coulomb barrier energy $E_\mathrm{C}$. To achieve this, similarly to \cite{somo98}, a Fermi-type function was used:
\begin{equation}
W=\frac{25 ~ \mathrm{MeV}}{1+e^{\left(0.9E_\mathrm{C}-E_\mathrm{c.m.} \right)/a_E}} \quad .
\end{equation}
The relevance of $0.9E_\mathrm{C}$ as the energy below which the potential parameters have to be strongly modified was pointed out in Ref. \cite{avri10}. The value $a_E=2$ MeV for the ``diffuseness'' of the Fermi-type function was varied but it was found that best agreement with the experimental data was achieved when using the same value as in Ref. \cite{somo98}. An often overlooked fact is that the choice of the Coulomb radius parameter $r_\mathrm{C}$ may have a strong impact. While the $S$ factors obtained with the OMPs by \cite{McFa66} and \cite{frohdip,raufroh} are quite insensitive to the chosen $r_\mathrm{C}$, the results obtained with the energy-dependent OMPs by \cite{somo98} and \cite{avri10} are strongly affected (see also \cite{Raus11} for a further discussion). Also, the results with our energy-dependent OMP are sensitive to $r_\mathrm{C}$ and we set $r_\mathrm{C}=1.2$ fm for a best fit, which is quite similar to the 1.25 fm used in Ref. \cite{somo98}.

The $S$ factors obtained with the fitted potential are shown in Fig. \ref{fig:fit} along with the experimental values. For comparison, the values obtained with the energy-independent OMP by \cite{McFa66} are also included again. Although not all of the detailed features seen in the data are reproduced, the overall energy-dependence of the $S$ factors is well described. (It is mentioned in passing that the two data at the lowest energies should be given lower weight in such a comparison regarding the $\alpha$ OMP because the $\alpha$ width is not the only uncertainty there.)

As explained in Secs.\ \ref{sec:introduction} and \ref{sec:sensitivities}, the relevant quantities in the $\gamma$ process are the ($\gamma$,$\alpha$) reaction rates [which, in turn, are calculated from ($\alpha$,$\gamma$) rates]. Therefore, it is interesting to examine the impact of our fitted OMP on $\alpha$ captures. A complete comparison as, for example, is performed in Ref. \cite{rau06}, is beyond the scope of this paper, but two cases can be discussed. Extending our OMP into the astrophysically relevant energy window for the reaction $^{141}$Pr($\alpha$,$\gamma$)$^{145}$Pm (see Fig.\ \ref{fig:Sensitivity_ag}) and calculating the stellar reaction rate for this reaction, it is found that the rate would be lower by three orders of magnitude compared to the one obtained with the potential of \cite{McFa66}. This is due to the very different energy dependence of the potential and the resulting cross sections. For comparison, the rate resulting from application of the OMP by \cite{frohdip,raufroh} is lower by only 10\% to 20\% in this case. The reduction in the $^{145}$Pm($\gamma$,$\alpha$) rate is the same as for the capture rate and thus it would become much slower than the competing $^{145}$Pm($\gamma$,n). This removes the $^{145}$Pm($\gamma$,$\alpha$) deflection point previously present in the $\gamma$-process path \cite{rau06}. Nevertheless, astrophysically this is of minor importance because there are no stable seed nuclei of Pm to be photodisintegrated and it could only change the processing of downflows from higher masses, which are small.

More relevant is the reaction $^{144}$Sm($\alpha$,$\gamma$)$^{148}$Gd, because it determines the $^{146}$Sm/$^{144}$Sm ratio in core-collapse supernova ejecta which can be determined via Nd/Sm ratios measured in meteoritic material \cite{somo98,arngorp,prinz,woosm144,rausm144}.
A calculation of the $S$ factors of $^{144}$Sm($\alpha$,$\gamma$)$^{148}$Gd with our potential showed that it yields values comparable at low energies to the fit potential in Ref. \cite{somo98} (see, however, Ref.\ \cite{Raus11} for a cautionary discussion of further uncertainties in that OMP). Due to the different energy dependence, however, it yields different rates in the relevant temperature range $2\leq T\leq 3$ GK. Compared with the calculation shown in Ref. \cite{somo98} the rate is higher by 50\% at 3 GK, which coincides with the rate given there at 2.5 GK, and is lower by a factor of 3 at 2 GK. Since the ejected $^{146}$Sm/$^{144}$Sm ratio is determined by photodisintegration at the lower end of the temperature range, this would result in an even larger $^{146}$Sm/$^{144}$Sm ratio as reported in Ref. \cite{somo98}. It would be very difficult to reconcile such a ratio with the ones found in meteorites. It is important to note that the energy dependence of our OMP parameters may be more complicated than assumed and the use of the potential at lower energies than the ones measured in this work may not be warranted.

\section{\label{sec:summary}Summary and Conclusion}

The cross sections of the reaction $^{141}$Pr($\alpha$,n)$^{144}$Pm were measured between 11 MeV and 15 MeV. The $\gamma \gamma$ coincidence method was used within one clover-type HPGe detector in an activation experiment to determine total cross sections. This method improved in this case the peak-to-background ratio in the coincidence spectra by a factor of more than 50 compared to the singles spectra, which minimize the sensitivity limit dramatically. Therefore, the $\gamma \gamma$ coincidence method allows measuring at lower energies close to the astrophysical relevant energy range. A comparison with 4 further detectors proves that the $\gamma \gamma$ coincidence method is an excellent tool to investigate cross sections down to the microbarn range.

It was possible to measure close to the astrophysically relevant energy range, allowing us to study the validity of $\alpha$+nucleus optical-model potentials at low energies. A comparison of the experimental results with Hauser-Feshbach statistical model calculations confirms that the theoretical description of sub-Coulomb $\alpha$ transitions remains problematic. Especially the reproduction of the energy dependence of the excitation function by the theoretical models is worse than in previously studied comparable cases. This may be connected to the closed neutron shell of $^{141}$Pr because strong discrepancies between experiment and predictions were also found for the reaction $^{144}$Sm($\alpha$,$\gamma$)$^{148}$Gd in a previous study.

A local energy-dependent optical potential was derived to improve the description of the present data, thereby proving the possibility to do so by just employing a suitable potential. It was further applied to calculate the astrophysical reaction rates for $^{141}$Pr($\alpha$,$\gamma$), $^{144}$Sm($\alpha$,$\gamma$), and their inverses. The revised rates would remove a $\gamma$-process path deflection at $^{145}$Pm and increase the $^{146}$Sm/$^{144}$Sm ratio produced in core-collapse supernovae, respectively. The energy-dependence of the potential, however, is still unconstrained in the energy range actually relevant for the astrophysical $\gamma$ process and definitive conclusions are premature.

The results underline the difficulties to determine an $\alpha$+nucleus OMP which is globally applicable not only for different target nuclei but also at astrophysically relevant, low interaction energies. This also underlines the importance of measuring as close as possible or resonable to the astrophysically required energies. In our case we decided not to measure at even lower energies because here the cross section is also sensitive to the neutron and $\gamma$ width and a determination of the $\alpha$+$^{141}$Pr OMP is challenging. In general, further investigations are necessary and global improvements cannot be expected before additional low-energy data for a wider range of targets are accumulated.

\begin{acknowledgments}
The authors acknowledge the help of the PTB accelerator staff participating in this experiment and M. Schmiedel from the activity department of the PTB. Moreover, we thank I. Dillmann and V. Derya for fruitful discussions during the preparation of the experiment and the data analysis. Many thanks to N. Warr for his helpful corrections. We thank P. Mohr for his valuable comments on the manuscript. ME, AH, and AS are members of the Bonn-Cologne Graduate School of Physics and Astronomy. This project has been supported by the Deutsche Forschungsgemeinschaft under the contract ZI 510/5-1. TR is supported by the European Commission within the FP7 ENSAR/THEXO project.

\end{acknowledgments}

%

\end{document}